\documentclass[aps,prc,twocolumn,showpacs,amsmath,amssymb,nofootinbib]{revtex4-2}
\usepackage{graphicx}
\usepackage{color}
\usepackage{times}
\usepackage{inputenc}
\usepackage{bm}
\usepackage{ulem}
\usepackage{multirow}
\usepackage{float}
\usepackage{url}
\usepackage{natbib}
\usepackage{mathrsfs}
\usepackage{physics}
\usepackage{comment}
\usepackage[table,xcdraw]{xcolor}
\usepackage{booktabs}
\usepackage{comment}
\usepackage[caption=false]{subfig}
\usepackage[colorlinks=true,citecolor=blue,urlcolor=blue,linkcolor=blue]{hyperref}
\newcommand{\ols}[1]{\mskip.4\thinmuskip\overline{\mskip-.4\thinmuskip {#1} \mskip-.4\thinmuskip}\mskip.4\thinmuskip} % overline short
\begin{document}
%%%%%%%%%%%%%%%%
%%%%%%%%%%%%%%%%%%%%%%%%%%%%%%%%%%%%%%%%%%%%%%%%%%%%%%
\title{Influence of Dark Matter on the Magnetized Neutron Star}
%%%%%%%%%%%%%%%%%%%%%%%%%%%%%%%%%%%%%%%%%%%%%%%%%%%%%%
%%%%%%%%%%%%%%%%%%%%%%%%%%%%
\author{Vishal Parmar$^{1}$}
\email{physics.vishal01@gmail.com}
\author{H. C. Das$^{2}$}
\email{harish.d@iopb.res.in}
\author{M. K. Sharma$^{1}$}
\author{S. K. Patra$^{3,4}$} 
%%%%%%%%%%%%%%%%%%%%%%%%%%%
\affiliation{\it $^{1}$ School of Physics and Materials Science, Thapar Institute of Engineering and Technology, Patiala 147004, India}
\affiliation{\it $^{2}$ INFN Sezione di Catania, Dipartimento di Fisica,
Via S. Sofia 64, 95123 Catania, Italy}
\affiliation{\it $^{3}$Institute of Physics, Sachivalaya Marg, Bhubaneswar 751005, India}
\affiliation{\it $^{4}$Homi Bhabha National Institute, Training School Complex, Anushakti Nagar, Mumbai 400094, India}
%%%%%%%%%%%%%
\date{\today}
%%%%%%%%%%%%%
%%%%%%%%%%%%%%%%
\begin{abstract}
%%%%%%%%%%%%%%%%
Over the past two decades, significant strides have been made in the study of Dark Matter (DM) admixed neutron stars and their associated properties. However, an intriguing facet regarding the effect of DM on magnetized neutron stars still remains unexplored. This study is carried out to analyze the properties of DM admixed magnetized neutron stars. The equation of state for the DM admixed neutron star is calculated using the relativistic mean-field model with the inclusion of a density-dependent magnetic field. Several macroscopic properties, such as mass, radius, particle fractions, tidal deformability, and the $f$-mode frequency, are calculated with different magnetic field strengths and DM configurations. The equation of state is softer with the presence of DM as well as for the parallel components of the magnetic field and vice-versa for the perpendicular one. Other macroscopic properties, such as mass, radius, tidal deformability, etc., are also affected by both DM and magnetic fields. The change in the magnitude of different neutron star observables is proportional to the amount of DM percentage and the strength of the magnetic field. We observe that the change is seen mainly in the core part of the star without affecting the crustal properties. 
%Therefore, the gross properties are changed due to both interactions present.
%%%%%%%%%%%%%%%%
\end{abstract}
%%%%%%%%%%%%%%%%
%%%%%%%%%%
\maketitle
%%%%%%%%%%
%%%%%%%%%%%%%%%%%%%%%%
\section{Introduction}
\label{intro}
%%%%%%%%%%%%%%%%%%%%%%
In recent years, the investigation of magnetars and pulsars, characterized as highly magnetized neutron stars, has emerged as a fascinating research field at the juncture of nuclear physics and astrophysics. These enigmatic celestial objects exhibit magnetic fields of remarkable strength ($B \sim \ 10^{17}-10^{18}$ G), surpassing those typically observed in neutron stars by several orders of magnitude \cite{Makishima_2014}. Such immensely powerful field conditions are presently beyond the reach of terrestrial laboratories. Consequently, pulsars and magnetars serve as extraterrestrial laboratories for examining and advancing physical theories. These objects present a plethora of exhilarating phenomena, including manifestations of exotic QED mechanisms like photon splitting and magnetic pair creation \cite{Baring_2001}, outburst and quiescent emissions \cite{Ellison_2000}, seismic activity \cite{Lander_2023}, dissipative processes in the magnetospheres \cite{Hu_2022}, axion-like particles \cite{Fortin_2021}, and dense matter physics \cite{Ferreira_2023}, among others. The exploration of  these physical phenomena establishes the study of magnetized neutron stars as a pivotal research area in astrophysics, offering valuable insights into the behaviour of matter and radiation in extreme environments \cite{Kaspi_2017}.

It is a well-known fact with compelling evidence that most of the matter in the Universe is dark matter (DM) \cite{BERTONE2005279}. Since neutron stars are highly compact and dense, the collision between the DM particles and constituents of the neutron star results in the loss of energy for the DM to become bound to the gravitational pull of the neutron star. Therefore, neutron stars have long been used as a tool in the quest to uncover the particle nature of DM and their scattering cross sections \cite{Bertone_2008, Nicole_2021, Ben_2021, Rafiei_2022}. The DM admixture neutron star results in significant deviation in the neutron star observables such as mass-radius profile, tidal deformation, luminosity \cite{Kouvaris_2008}, accretion \cite{Lavallaz_2010}, etc.,  and hence, can act as a probe to measure the DM properties indirectly. 

In the last two decades, there have been numerous attempts to study the DM admixed neutron star and associated properties \cite{Sandin_2009, Ciarcelluti_2011, Ellis_2018, Hooper_2004, Bertone_2008, Kouvaris_2008, Das_2019, Das_2021, Das_2022, Das_fmode_PRD_2021, DasMNRAS_2021, Routaray2023May, Lopes2023May}. However, despite significant progress, the impact of DM on magnetized neutron stars remains a scientific question that necessitates immediate exploration. Since most of the observed neutron stars are either pulsars or magnetars \cite{Manchester_2005}, it becomes essential to study the influence of DM on these compact stars. Recently strange star admixed with fermionic DM in a strong magnetic field (MF) was analyzed using the MIT bag model \cite{Ferreira_2023}. It was shown that tidal deformability gets  intensely affected by such stars. 

In this paper, we aim to present an analysis of the unexplored influence of DM on pure hadronic magnetized neutron stars over a MF range ($10^{17}-10^{19}$G). We use the magnetized neutron star formalism as described in \cite{Broderick_2000, Strickland_2012, Parmar_2023} and the effective relativistic mean field (E-RMF) model for the nuclear interaction. The E-RMF theory has been successfully applied to a wide range of nuclear physics problems ranging from finite nuclei to the neutron star \cite{Patra_2002, Kumar_2018, Ankit_2020, Das_fmode_PRD_2021, Parmar_2022_1}. This theory has recently been used to address the DM admixed neutron star \cite{Hcdas_2021, Hcdas_2020}. In the present work, we use two E-RMF models, namely, BigApple \cite{Fattoyev_2020} and IOPB \cite{Kumar_2018}. The BigApple force is designed to account for the $2.6 M_\odot$ and original constraint on the tidal deformability of a $1.4 M_\odot$ neutron star in accordance with the secondary component of GW190814 \cite{Abbott_2020}. At the same time, the IOPB parameter set reproduces the maximum mass from massive pulsars such as PSR J0740+6620, which estimate that the neutron star mass should be greater than $2~M_\odot$ ($M = 2.14_{-0.09}^{+0.10} \ M_\odot$) \cite{Cromartie_2020}). These parameter sets also reproduce the nuclear matter and finite nuclei observables in agreement with empirical and observational constraints \cite{Kumar_2018, DasBig_2020}.

To model the DM interaction with the neutron star, we consider the Neutralino as a DM candidate, which belongs to WIMP as taken in Ref. \cite{Panotopoulos_2017}. Further, DM is treated analogous to a neutron, considering the DM as a charge-less fermion. Other types of DM candidates are also hypothesized, such as Bosonic, asymmetric, etc., having different properties compared to the Fermionic one. Several works have been done to explore such types of scenarios that explore the effects of DM on the different NS properties \cite{Panotopoulos_2017, Das_2019, Das_2021, Das_2022, DasMNRAS_2021, Das_fmode_PRD_2021, Routaray2023May}. However, there are still windows to explore the DM effects on other properties of the compact star, its particle nature, etc.

Henceforth, we aim to investigate the possible changes in the DM admixed magnetized neutron star equation of state (EOS), its composition, and neutron star observables, which include mass-radius relations, tidal deformability, and $f$-mode oscillation.  Although the neutron star should be deformed due to the anisotropic pressure in the presence of the MF, for our purposes, we will adopt, as a preliminary assumption, that the configuration of a highly magnetized neutron star can be adequately characterized by utilizing the conventional spherically symmetric equations governing stellar structure. These assumptions are based on the fact that i) the monopole nature of the chaotic magnetic field inside the star validates the use of the spherical symmetry for the background metric \cite{flores2020gravitational}, ii) in order to construct realistic models of magnetized neutron stars, it is crucial to incorporate both poloidal and toroidal MF components simultaneously \cite{Frieben_2012, Ciolfi_2013}. It has been established that exclusively toroidal MFs cause the NS to become prolate, while purely poloidal MFs tend to render it oblate in shape. However, when the toroidal and poloidal components are comparable in magnitude, it is plausible to anticipate that the oblateness and prolateness effects approximately cancel out, resulting in stars that closely approximate spherical symmetry \cite{Mariani_2019}. This assumption holds good for the range of MF $10^{15}-10^{18}$ as the deformation from spherical symmetry turns out to be less than 1\% \cite{Patra_2020, Bordbar2022, Chu_2015}. 
It is seen that with an increasing DM mass, the maximum mass of the neutron star decreases \cite{Das_2019}. On the other hand, it was shown that the maximum mass of the neutron star is an increasing/decreasing function of the MF depending on the perpendicular/parallel pressure \cite{Patra_2020, Rather_2021, Bordbar2022}. Therefore, it is interesting to examine the combined effect of DM admixed neutron star. Moreover, in the present work, we present separate results for the perpendicular and parallel pressure in line with Refs. \cite{Chu_2015, Huang_2010, Patra_2020}.

The paper's organization is as follows: In Sec \ref{formulation}, we describe the effect of MF on the EOS employing the E-RMF framework and DM Model. The results indicating the EOS, composition, mass-radius relations, tidal deformability, and $f$-mode oscillation are discussed in Sec. \ref{results}. Finally, we summarize our results in Sec. \ref{conclusion}.
%%%%%%%%%%%%%%%%%%%
\section{Formalism}
\label{formulation} 
%%%%%%%%%%%%%%%%%%%
%%%%%%%%%%%%%%%%%%%%%%
\subsection{RMF model}
%%%%%%%%%%%%%%%%%%%%%%
The effective Lagrangian in the E-RMF, which include  $\sigma$, $\omega$, $\rho $, $\delta$, and photon in association with the baryons  can be written as  \cite{Patra_2002, MULLER_1996, Wang_2000, Kumar_2020, Das_2021},
%%%%%%%%%%%%%%%%
\begin{widetext}
\begin{eqnarray}
\label{rmftlagrangian}
\mathcal{E}(r)&=&\psi^{\dagger}(r)\qty{i\alpha\cdot\grad+\beta[M-\Phi(r)-\tau_3D(r)]+W(r)+\frac{1}{2}\tau_3R(r)+\frac{1+\tau_3}{2} A(r)-\frac{i\beta \alpha }{2M}\qty(f_\omega \grad W(r)+\frac{1}{2}f_\rho \tau_3 \grad R(r))}\psi(r) \nonumber \\
&+& \qty(\frac{1}{2}+\frac{k_3\Phi(r)}{3!M}+\frac{k_4}{4!}\frac{\Phi^2(r)}{M^2})\frac{m^2_s}{g^2_s}\Phi(r)^2+\frac{1}{2g^2_s}\qty\Big(1+\alpha_1\frac{\Phi(r)}{M})(\grad \Phi(r))^2-\frac{1}{2g^2_\omega}\qty\Big(1+\alpha_2\frac{\Phi(r)}{M})(\grad W(r))^2 \nonumber\\
&-&\frac{1}{2}\qty\Big(1+\eta_1\frac{\Phi(r)}{M}+\frac{\eta_2}{2}\frac{\Phi^2(r)}{M^2})\frac{m^2_\omega}{g^2_\omega}W^2(r)-\frac{1}{2q^2}(\grad A^2(r))^2 -\frac{1}{2g^2_\rho}(\grad R(r))^2
-\frac{1}{2}\qty\Big(1+\eta_\rho\frac{\Phi(r)}{M})\frac{m^2_\rho}{g^2_\rho}R^2(r)\nonumber \\
&-&\frac{\zeta_0}{4!}\frac{1}{g^2_\omega}W(r)^4-\Lambda_\omega(R^2(r)W^2(r))
+\frac{1}{2g^2_\delta}(\grad D(r))^2
+\frac{1}{2}\frac{m^2_\delta}{g^2_\delta}(D(r))^2.
\end{eqnarray} 
\end{widetext}
%%%%%%%%%%%%%%
Here $\Phi(r)$, $W(r), R(r), D(r)$ and $A(r)$ are the fields corresponding to $\sigma$, $\omega$, $\rho$ and $\delta $ mesons and photon, respectively. The $g_s$, $g_{\omega}$, $g_{\rho}$, $g_{\delta}$ and $\frac{q^2}{4\pi }$ ($q=e$) are the corresponding coupling constants and $m_s$, $m_{\omega}$, $m_{\rho}$ and $m_{\delta}$ are the corresponding masses. M is the mass of the nucleon.
 
In the process of fitting, the coupling constants of the effective Lagrangian are ascertained using a collection of experimental data that considers a significant portion of the vacuum polarization effects within the no-sea approximation which is essential to determine the stationary solutions of the relativistic mean-field equations \cite{Kumar_2018, Reinhard_1989}. The effective masses of proton, $M^*_p$, and neutron, $M^*_n$, are written as

\begin{eqnarray}
\label{eq:effmass}
    M^*_p&=M-\Phi(r)-D(r)\\
    M^*_n&=M-\Phi(r)+D(r)
\end{eqnarray}

Finally, the zeroth component $T_{00}= H$ and the third component $T_{ii}$ of energy-momentum tensor \cite{Serot_1986}
%%%%%%%%%%%%%%%%
\begin{equation}
\label{set}
T_{\mu\nu}=\partial_\nu\phi(x)\frac{\partial\mathcal{E}}{\partial(\partial^\mu \phi(x))}-g_{\nu\mu}\mathcal{E},
\end{equation}
%%%%%%%%%%%%%%
yields the energy density and pressure, respectively \cite{Kumar_2018}.   The expressions for energy density, pressure, and mean-field equations can be found in Ref. \cite{Kumar_2018}.

\subsection{DM Model}
%%%%%%%%%%%%%%%%%%%%%
In this section, we provide the formalism for the DM admixed neutron star.  The DM particles are captured inside the NSs due to their huge gravitational potential and immense baryon density. In this study, we choose the single fluid model, where the DM particles interact with nucleons by exchanging the Higgs. Therefore, the system energy density and pressure are the addition of both nuclear matter and DM. The DM particles are accreted inside the NS mainly in the core part due to higher gravitational potential in comparison to the crust \cite{Panotopoulos_2017, Das_2019, Das_2021}. Hence, in single-fluid approximation, only one scenario is seen that the DM are uniformly distributed throughout the core part, which is independent of any fractions inside the star. 
However, in the case of the two-fluid model, there are different scenarios seen, such as DM core and DM halo, due to no interactions with nucleons rather than self gravitational interaction between DM. That means the evolution mechanism is completely different. The magnitude of interactions mainly depends on the type of DM particle and its fraction inside the star. For more details, see the references \cite{Nelson_2019, Cassing_2023}.

In this study, we choose the single-fluid model. Therefore, to calculate the DM fractions, we assumed that the DM density is uniformly distributed inside the neutron star \cite{Panotopulos_2017, Das_2019, Das_2021}.
Assuming the average number density of nucleons ($\rho_b$) is $10^3$ times larger than the average DM density ($\rho_{\rm DM}$), which implies the ratio of the DM and the neutron star mass to be $\sim \frac{1}{6}$ \cite{Panotopoulos_2017}. 
Since the nuclear saturation density is $\rho_0  \sim 0.16$ fm$^{-3}$, therefore, the DM number density becomes $\rho_{\rm DM} \sim 10^{-3}\rho_0\sim 0.16\times 10^{-3}$ fm$^{-3}$. Using the $\rho_{\rm DM}$, the $k_{f}^{\rm DM}$ is obtained from the equation $k_f^{\rm DM}=(3\pi^2 \rho_{\rm DM})^{1/3}$. Hence the value of $k_{f}^{\rm DM}$ is $\sim 0.033$ GeV. Therefore, in our case, we vary the DM momenta from 0, 0.02, and 0.04 GeV. In this study, we choose Neutralino as a DM candidate, which belongs to WIMPS. The interacting Lagrangian is in the following ~\cite{Panotopoulos_2017, Das_2019, Hcdas_2020, Das_2021, DasMNRAS_2021}:
%%%%%%%%%%%%%%%%
\begin{eqnarray}
{\cal{L}}_{\rm DM}&=&  \bar \chi \Big[ i \gamma^\mu \partial_\mu - M_\chi + y h \Big] \chi +  \frac{1}{2}\partial_\mu h \partial^\mu h 
\nonumber\\
&-& \frac{1}{2} M_h^2 h^2 +\frac{fM}{v} \bar{\psi} h \psi , 
\label{eq:ldm}
\end{eqnarray}
%%%%%%%%%%%%%
$\psi$ and $\chi$ are the baryons and DM wave functions respectively. Here, we choose values of the DM-Higgs coupling ($y$), proton-Higgs form factor ($f$), and vacuum value ($v$) of Higgs are 0.07, 0.35, and 246 GeV, respectively, as considered in Refs. \cite{Hcdas_2020, Das_2021}. The free parameters are constrained with the help of DM detection data available to date \cite{Das_2021}. With this preliminary information, we can calculate the DM scalar density ($\rho_s^{\rm DM}$), energy density, and pressure using the mean-field approximation as done in Refs. \cite{Das_2019, Panotopoulos_2017}
%%%%%%%%%%%%%%%%
\begin{equation}
\rho_s^{\rm DM}=\langle\bar{\chi}\chi\rangle =  \frac{\gamma}{2 \pi^2}\int_0^{k_f^{\rm DM}} k^2\ dk \  \frac{M_\chi^\star}{\sqrt{M_\chi^\star{^2}+ k^2}},
\label{eq:dm_density}
\end{equation}
%%%%%%%%%%%%%%
where $k_f^{\rm DM}$ is the Fermi momentum for DM. $\gamma$ is the spin degeneracy factor with a value of 2 for neutron and proton. 

The energy density (${\cal{E}}_{\rm DM}$) and pressure ($P_{\rm DM}$) for neutron star with DM can be obtained by solving the Eq. (\ref{eq:ldm})
\begin{eqnarray}
\epsilon_{\rm DM}& = & \frac{1}{\pi^2}\int_0^{k_f^{\rm DM}} k^2 \ dk \sqrt{k^2 + (M_\chi^\star)^2 } +\frac{1}{2}M_h^2 h_0^2 ,
\label{eq:edm}
\end{eqnarray}
and
\begin{eqnarray}
P_{\rm DM}& = & \frac{1}{3\pi^2}\int_0^{k_f^{\rm DM}} \frac{ k^4 \ dk} {\sqrt{k^2 + (M_\chi^\star)^2}} - \frac{1}{2}M_h^2 h_0^2 ,
\label{eq:pdm}
\end{eqnarray} 
$M_h$ is the Higgs mass equal to 125 GeV, and $h_0$ is the Higgs field.  With the Higgs contribution, the effective mass of the system becomes
%%%%%%%%%%%%%%%%
\begin{equation}
\label{eq:eff_mass}
    M^*_{p,n}= M-\Phi \mp D(r) - \frac{fM}{v}h_0.
\end{equation}
%%%%%%%%%%%%%%%

%%%%%%%%%%%%%%%%%%%%%%%%%%%%%%%%

\subsection{Effects of magnetic field}
 
In the presence of a uniform external magnetic field
pointing in the $z$ direction ($B=B\hat{z}$) such that $\vec{\nabla}\cdot \vec{B}=0$ \cite{fang_2017}, the transverse momenta of the charged
particles with an electric charge $q$ are restricted to discrete
Landau levels \cite{Strickland_2012}. One can define the thermodynamic potential $\Omega$ \cite{Ashok_2001}, which depends upon the chemical potential ($\mu$), temperature (T) and magnetic field ($B$) such that it follows canonical relations $\Omega= -P_\parallel=\epsilon-\sum_i\rho_i\mu_i$ and $P_\perp= P_\parallel-MB$, with $\epsilon$ being the energy density, $\rho_i$ the density of $i^{th}$ particle, $\mu_i$ the corresponding chemical potential, $M$  the magnetisation of the system, $P_\parallel$ and $P_\perp$ are the pressure in the parallel and the transverse to the magnetic field direction \cite{Patra_2020, Strickland_2012, Ashok_2001, Broderick_2000}. A detailed description and derivations of the various quantities required to define the magnetised nuclear matter can be found in Refs. \cite{Broderick_2000, Strickland_2012, Ashok_2001}. Here, we present the necessary formalism required in the zero-temperature limit following Refs. \cite{Broderick_2000, Strickland_2012}.

The energy spectrum of the proton, which gets modified due to the Landau level, is written as \cite{Broderick_2000, Strickland_2012}
%%%%%%%%%%%%%%%%
\begin{equation}
    E_p=\sqrt{k_z^2+\ols{M}_{n,\sigma_z}^{p^2}}+W-R/2,
\end{equation}
%%%%%%%%%%%%%%%
and for charged leptons (electron and muon) as
%%%%%%%%%%%%%%%%
\begin{equation}
    E_{e,\mu}=\sqrt{k_z^2+\ols{M}_{n,\sigma_z}^{{e,\mu}^2}},
\end{equation}
%%%%%%%%%%%%%%%
where 
%%%%%%%%%%%%%%
\begin{align}
        \ols{M}_{n,\sigma_z}^{p^2}=M_{p}^{*^2}+2\Big(n+\frac{1}{2}-\frac{1}{2}\frac{q}{|q|}\sigma_z \Big)|q|B.\\
        \ols{M}_{n,\sigma_z}^{(e,\mu)^2}=M_{(e,\mu)}^{2}+2\Big(n+\frac{1}{2}-\frac{1}{2}\frac{q}{|q|}\sigma_z \Big)|q|B.
\end{align}
%%%%%%%%%%%%%
Here, $\sigma_z$ is the spin along the axis of the MF ($B$), $n$ is the principal quantum number, and $k_z$ is the momentum along the direction of the MF. $M^*_{p}$ is the effective mass for the proton defined in Eq. \eqref{eq:effmass}. The neutron spectrum is similar to the Dirac particle and takes form,
%%%%%%%%%%%%%%%%
\begin{equation}
    E_n=\sqrt{k^2+M_n^{*^2}}+W+R/2.
\end{equation}
%%%%%%%%%%%%%%%
At $T=0$, the number and energy density at zero temperature and in the presence of a MF is given by \cite{Broderick_2000}
%%%%%%%%%%%%%%%%
\begin{equation}
\label{eq:density}
    \rho_{i=e,\mu,p}=\frac{|q|B}{2 \pi^2} \sum_{\sigma_z} \sum_{n=0}^{n_{max}}k_{f,n,\sigma_z}^{i},
\end{equation}
%%%%%%%%%%%%%%%
\begin{eqnarray}
\label{eq:energy}
    \epsilon_{i=e,\mu,p}&=&\frac{|q|B}{4 \pi^2} \sum_{\sigma_z} \sum_{n=0}^{n_{max}} \nonumber\\
    &\cross& \Big[E_f^{i}k_{f,n,\sigma_z}^{i}     + \ols{M}_{n,\sigma_z}^{{i^2}} \ln \Big(\Big|\frac{E_f^i+k_{f,n,\sigma_z}^i}{\ols{M}_{n,\sigma_z}^{{i}}}\Big|\Big)\Big].
\end{eqnarray}
%%%%%%%%%%%%%%%
respectively. In above equations, $k_{f,n,\sigma_z}^i$ is defined by
%%%%%%%%%%%%%%%%
\begin{equation}
\label{eq:fermi_momentum}
    k_{f,n,\sigma_z}^{i^2}=E_f^{i^2}-\ols{M}_{n,\sigma_z}^{{i^2}},
\end{equation}
%%%%%%%%%%%%%%%
where the Fermi energies are fixed by the respective chemical potentials given by
%%%%%%%%%%%%%
\begin{align}
    E_f^{l=e,\mu}=\mu_{\mu,e},\\
    E_f^{b=p,n}=\mu_b-W \pm R/2.
\end{align}
%%%%%%%%%%%%
In Eqs. (\ref{eq:density}) and (\ref{eq:energy}), the $n_{\rm max}$ is the integer for which the Fermi momentum remains positive in Eq. (\ref{eq:fermi_momentum}) and is written as
%%%%%%%%%%%%%
\begin{align}
\label{eq:nmax}
    &n_{\rm max}=\Bigg[\frac{E_f^{p^2}-M^{*^2}_p}{2|q|B}\Bigg], \; {\rm proton}\\ \nonumber
    &n_{\rm max}=\Bigg[\frac{E_f^{i^2}-M^2_i}{2|q|B}\Bigg], \; \; {\rm electron} \; \& \; {\rm muon} \, .
\end{align}
%%%%%%%%%%%%
Here $[x]$ represents the greatest integer less than or equal to $x$. The scalar density for the protons is further determined as  \cite{Broderick_2000}
%%%%%%%%%%%%%%%%
\begin{equation}
    \rho_p^s=\frac{|q|BM^*_p}{2\pi^2}\sum_{\sigma_z} \sum_{n=0}^{n_{max}}\ln \Big(\Big|\frac{E_f^p+k_{f,n,\sigma_z}^p}{\ols{M}_{n,\sigma_z}^{{p}}}\Big|\Big).
\end{equation}
%%%%%%%%%%%%%%
The number, scalar, and energy density for the neutrons are similar to the field-free case and can be written as \cite{Kumar_2018, Patra_2002}.
 \begin{align}
     \label{eq:nn}
     \rho_n&=\frac{k_f^{n^3}}{3 \pi^2},\\
      \rho_n^s&=\frac{M^*_n}{2 \pi^2} \Big[E_f^nk_f^n -M^{*^2}_n \ln\Big(\Big| \frac{E_f^n +k_f^n}{M^*_n}  \Big|\Big)\Big]\\
      \epsilon_n&=\frac{1}{8 \pi^2} \Big[2 E_f^{n^3}k_f^n -M^{*^2}_nE_f^nk^n_f-M^{*^4 }_n \ln \Big(\Big|\frac{E_f^n+k_{f}^n}{M^*_n}\Big|\Big)\Big]
 \end{align}
Similarly, since the DM are considered chargeless fermion, the scalar density and the energy density follow from Eqs. \eqref{eq:dm_density} and \eqref{eq:edm}. The total baryon energy density, which  is the sum of matter-energy density and the contribution from the electromagnetic field, is written as \cite{Kumar_2018, Parmar_2022_1, Das_2019}

\begin{eqnarray}
 \epsilon_b&=& \epsilon_p + \epsilon_n+ +\rho_bW +\frac{1}{2}\rho_{3}R 
\nonumber\\
&+&\frac{ m_{s}^2\Phi^{2}}{g_{s}^2}\Bigg(\frac{1}{2}+\frac{\kappa_{3}}{3!}
\frac{\Phi }{M} + \frac{\kappa_4}{4!}\frac{\Phi^2}{M ^2}\Bigg) -\frac{1}{4!}\frac{\zeta_{0}W^{4}}{g_{\omega}^2}
\nonumber\\
&-&\frac{1}{2}m_{\omega}^2\frac{W^{2}}{g_{\omega}^2}\Bigg(1+\eta_{1}\frac{\Phi}{M}+\frac{\eta_{2}}{2}\frac{\Phi ^2}{M^2}\Bigg)
\nonumber\\
&-&\Lambda_{\omega}  (R^{2}\times W^{2})-\frac{1}{2}\Bigg(1+\frac{\eta_{\rho}\Phi}{M}\Bigg)\frac{m_{\rho}^2}{g_{\rho}^2}R^{2}
\nonumber\\
&+&\frac{1}{2}\frac{m_{\delta}^2}{g_{\delta}^{2}}D^{2}+    \frac{B^2}{8 \pi},
\label{eq:totenergy}
\end{eqnarray}

The  pressure due to baryon then can be written as \cite{Kumar_2018, Das_2019, Parmar_2022_1}
\begin{eqnarray}
P_b( \parallel/\perp)&=&  P_p(\parallel/\perp) +P_n+\frac{1}{4!}\frac{\zeta_{0}W^{4}}{g_{\omega}^2}
\nonumber\\
&&
-\frac{ m_{s}^2\Phi^{2}}{g_{s}^2}\Bigg(\frac{1}{2}+\frac{\kappa_{3}}{3!}
\frac{\Phi }{M}+ \frac{\kappa_4}{4!}\frac{\Phi^2}{M^2}\Bigg)
\nonumber\\
&&
+\frac{1}{2}m_{\omega}^2\frac{W^{2}}{g_{\omega}^2}\Bigg(1+\eta_{1}\frac{\Phi}{M}+\frac{\eta_{2}}{2}\frac{\Phi ^2}{M^2}\Bigg)
\nonumber\\
&&
+\Lambda_{\omega} (R^{2}\times W^{2})+\frac{1}{2}\Bigg(1+\frac{\eta_{\rho}\Phi}{M}\Bigg)\frac{m_{\rho}^2}{g_{\rho}^2}R^{2}
\nonumber\\
&&
-\frac{1}{2}\frac{m_{\delta}^2}{g_{\delta}^{2}}D^{2} (-/+) \frac{B^2}{8 \pi}.
\label{eq:press}
\end{eqnarray}
%%%%%%%%%%%%%%%%%%
While the contribution of the neutrons to the pressure is straightforward \cite{Patra_2020, Kumar_2018}, the contribution from the protons can be written in terms of  parallel ($P_\parallel$) and perpendicular ($P_\perp$) component as \cite{Chakrabarty_1996, Strickland_2012}, 

%%%%%%%%%%%%%%%%
\begin{eqnarray}
    \label{eq:p_par}
    P_{(p,\parallel)}=\frac{|q|B}{4 \pi^2} \sum_{\sigma_z=\pm 1} \sum_{n=0}^{n=n_{max}} \Bigg[E_f^ik_{f,n,\sigma_z}^i- \ols{M}_{n,\sigma_z}^{{i^2}} \\ \nonumber
     \times \ln \Big(\Big|\frac{E_f^i+k_{f,n,\sigma_z}^i}{\ols{M}_{n,\sigma_z}^{{i}}}\Big|\Big) \Bigg].
\end{eqnarray}
%%%%%%%%%%%%%%

%%%%%%%%%%%%%%%%
\begin{eqnarray}
    \label{eq:trans_press}
    P_{(p,\perp)}=\frac{|q|^2B^2}{2 \pi^2} \sum_{\sigma_z=\pm 1} \sum_{n=0}^{n=n_{max}} n \ln \Big(\Big|\frac{E_f^i+k_{f,n,\sigma_z}^i}{\ols{M}_{n,\sigma_z}^{{i}}}\Big|\Big) .
\end{eqnarray}
%%%%%%%%%%%%%%
In order to estimate the neutron star EoS, the $\beta-$ equilibrium under weak interaction and charge neutrality  constraints the population of various particles as 
%%%%}
\begin{subequations}
\begin{equation}
\mu_n=\mu_p+\mu_e,  \ \ \mu_e=\mu_\mu.
\end{equation}
\begin{equation}
    \rho_p=\rho_e+\rho_\mu,
\end{equation}
\end{subequations}
%%%%
where $\mu_{p,n,e,\mu}$ are the chemical potential of the proton, neutron electron, and muon, respectively. Finally, the total energy and pressure of the Dark Matter admixed Magnetized neutron star can be written as

\begin{eqnarray}
\epsilon=\epsilon_b+\epsilon_{DM}+\sum_{l=e,\mu}\epsilon_l\\
    P=P_b+P_{DM}+\sum_{l=e,\mu}P_l
\end{eqnarray}
It is relevant to mention here that in the literature, B-dependent (divergent) vacuum contribution was also accounted for when investigating the quark system to explain phenomena such as Magnetic catalysis \cite{Ebert_1999, Ferrari_2012, Fraga_2012}. Such an attempt, using the  relativistic mean-field description of nuclear matter considering baryons, was carried out by Ref. \cite{Haber_2014} and more recently by Ref. \cite{Patra_2020}.  While it was observed that the influence of vacuum contribution on the binding energy of symmetric nuclear matter becomes significant only under very strong magnetic fields \cite{Haber_2014}, our results remain unaffected by the utilization of B-dependent vacuum contribution.

%%%%%%%%%%%%%%%
\begin{figure*}
    \centering
    \includegraphics[width=0.55\textwidth]{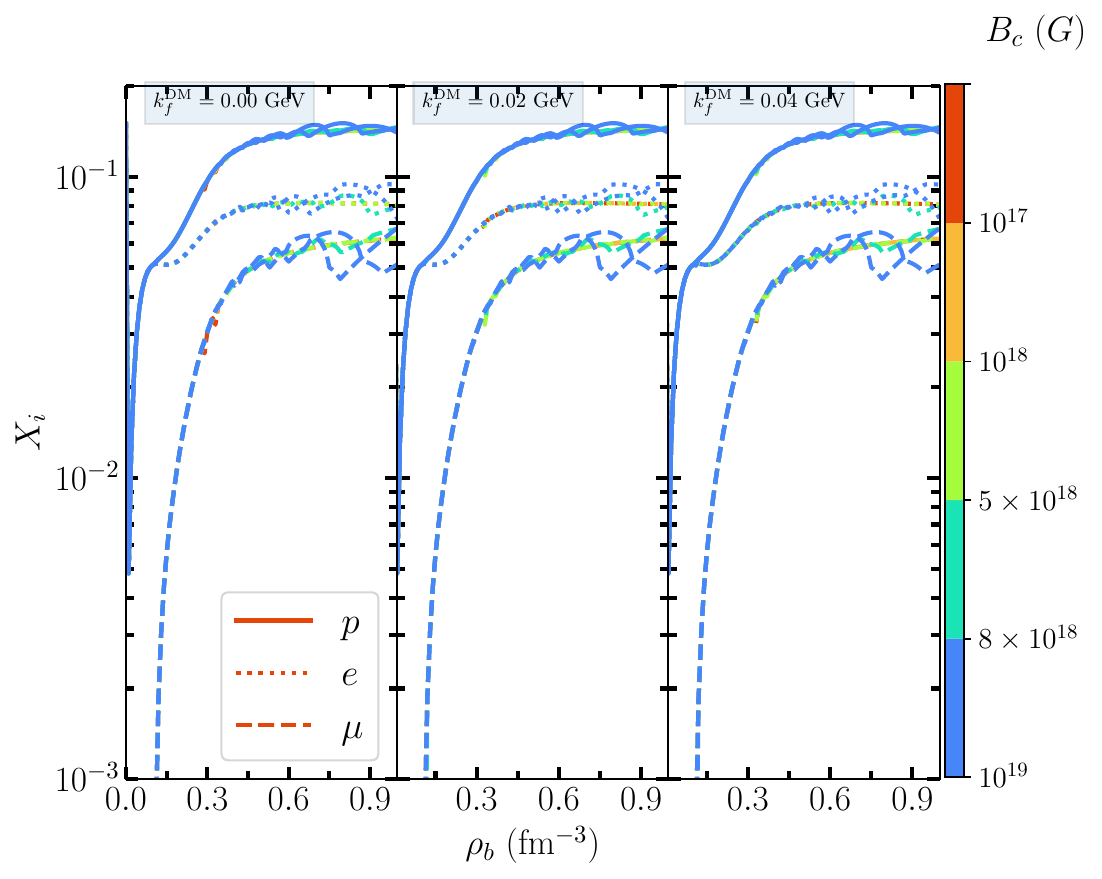}%%
    \includegraphics[width=0.55\textwidth]{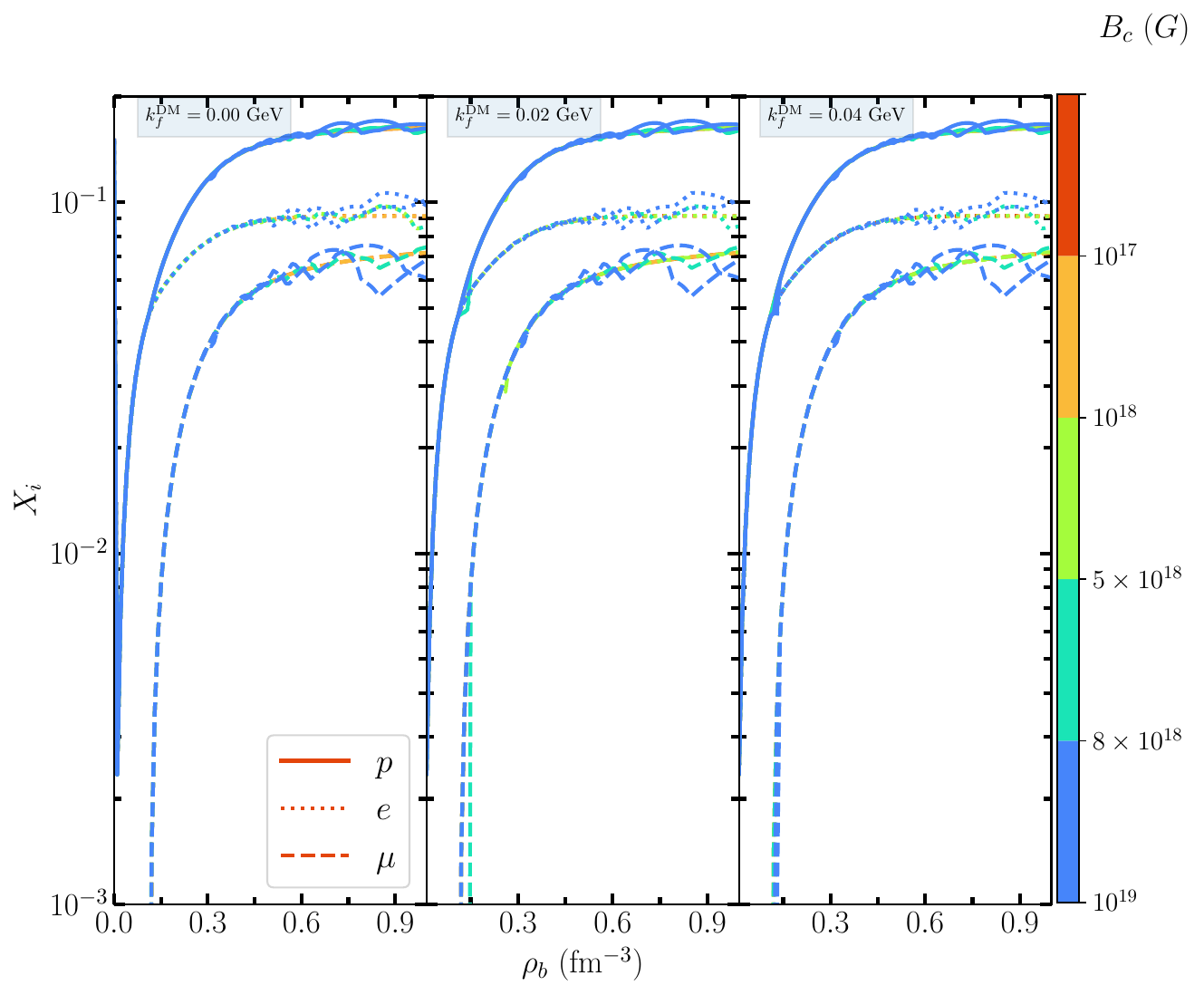}
    \includegraphics[width=0.55\textwidth]{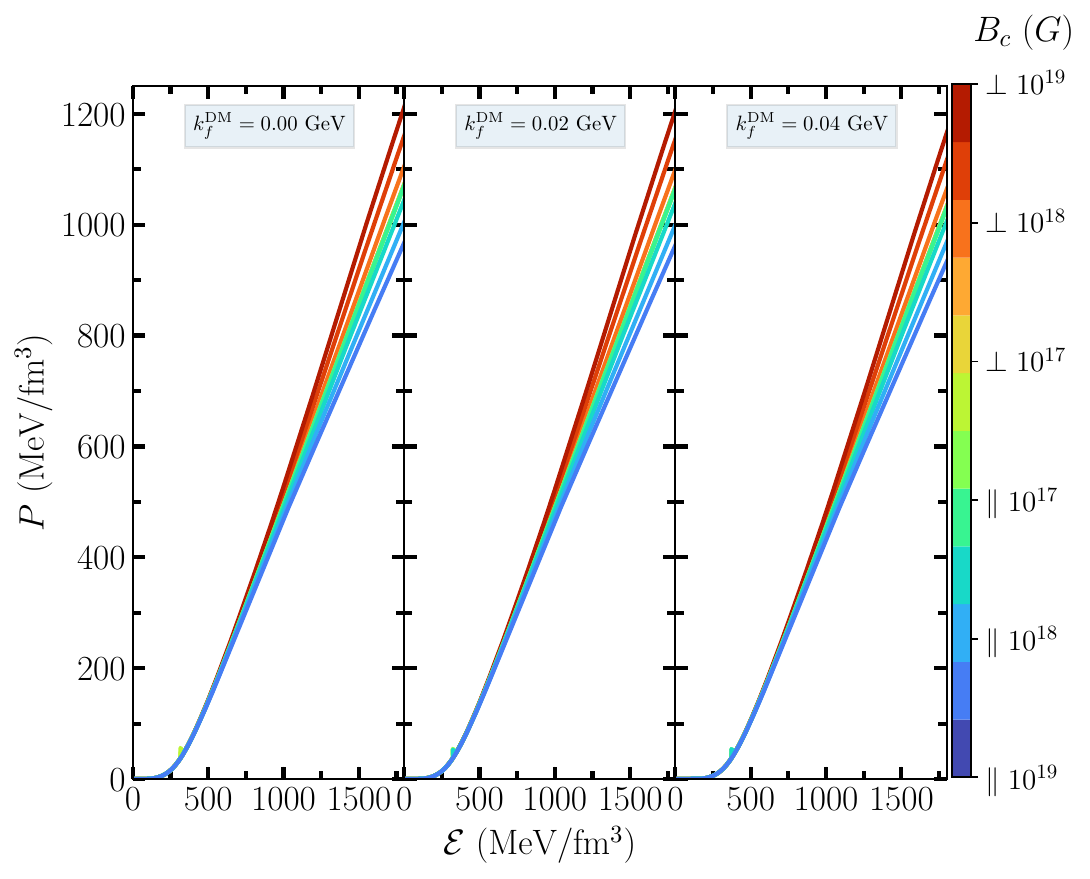}%%
    \includegraphics[width=0.55\textwidth]{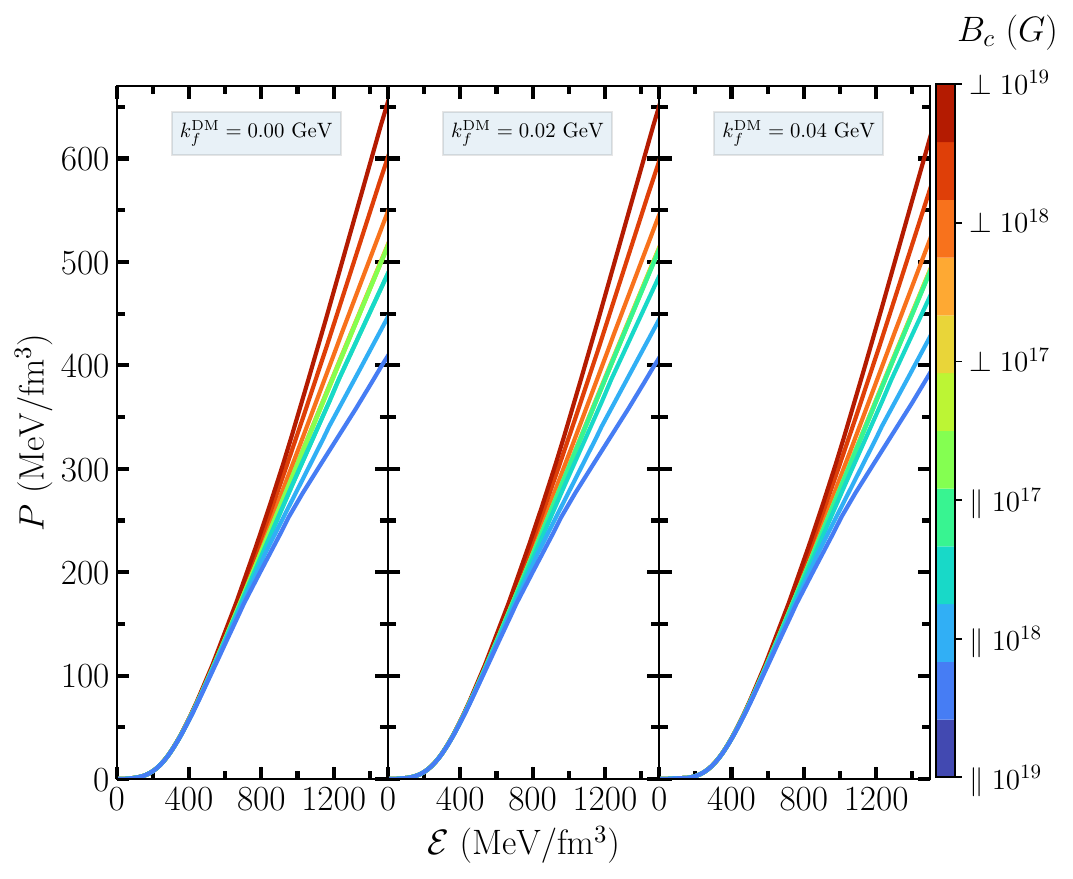}
    \caption{The particle fractions  (upper panel)  of different species with the presence of MF and DM having Fermi momenta 0.00, 0.02, and 0.04 GeV, respectively for BigApple (left panel), and IOPB-I (right panel) parameter sets and their  EoSs (lower panel). }
    \label{fig:eos_par_frac}
\end{figure*}
%%%%%%%%%%%%%%
%%%%%%%%%%%%%%%%%%%%%%%%%%%%%%%%%%%%%%%%%%%%%
\subsection{Density-dependent MF}
%%%%%%%%%%%%%%%%%%%%%%%%%%%%%%%%%%%%%%%%%%%%%
In this work, the MF ($B$) is parametrized from the surface to the center of the star as 
\cite{Rabhi_2009, Debades_1997, Mallick_2014, Bordbar2022}
%%%%%%%%%%%%%%%%
\begin{equation}
    B\Bigg(\frac{\rho}{\rho_0}\Bigg)=B_{\rm surf}+B_c\Bigg( 1- \exp{-\beta\Bigg(\frac{\rho}{\rho_0}\Bigg)^\gamma}\Bigg). 
\end{equation}
%%%%%%%%%%%%%%
Here, $\rho_0$ is the saturation density, $B_{\rm surf}$ is the surface MF taken to be $10^{15}$ G and $B_c$ is the MF at the center of the star. The parameters $\beta=0.02$ and $\gamma=3.00$ are chosen to reproduce the observational MF \cite{Casali_2014}. 
%%%%%%%%%%%%%%%%%%%%%

\section{Results and Discussion}
\label{results} 
%%%%%%%%%%%%%%%%%%%%%%%%%%%%%%%%
%%%%%%%%%%%%%%%
\begin{figure*}
    \centering
    \includegraphics[width=0.55\textwidth]{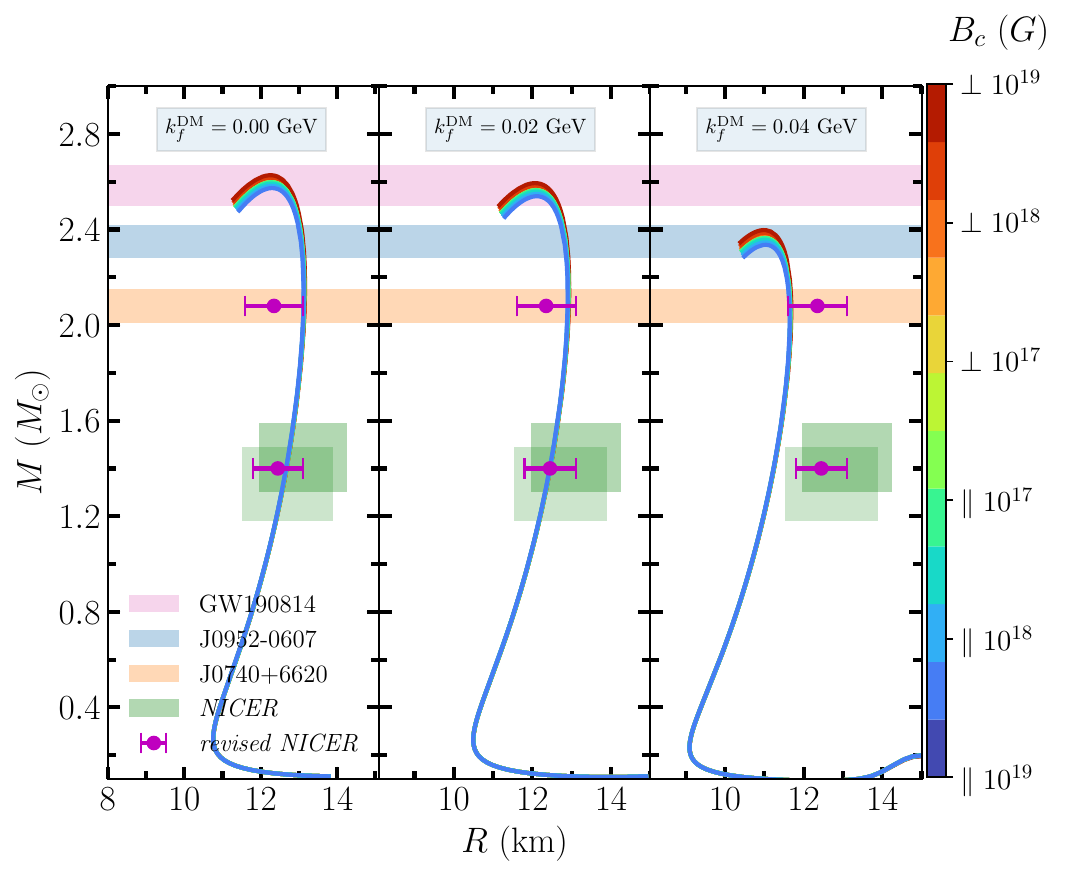}%%
    \includegraphics[width=0.55\textwidth]{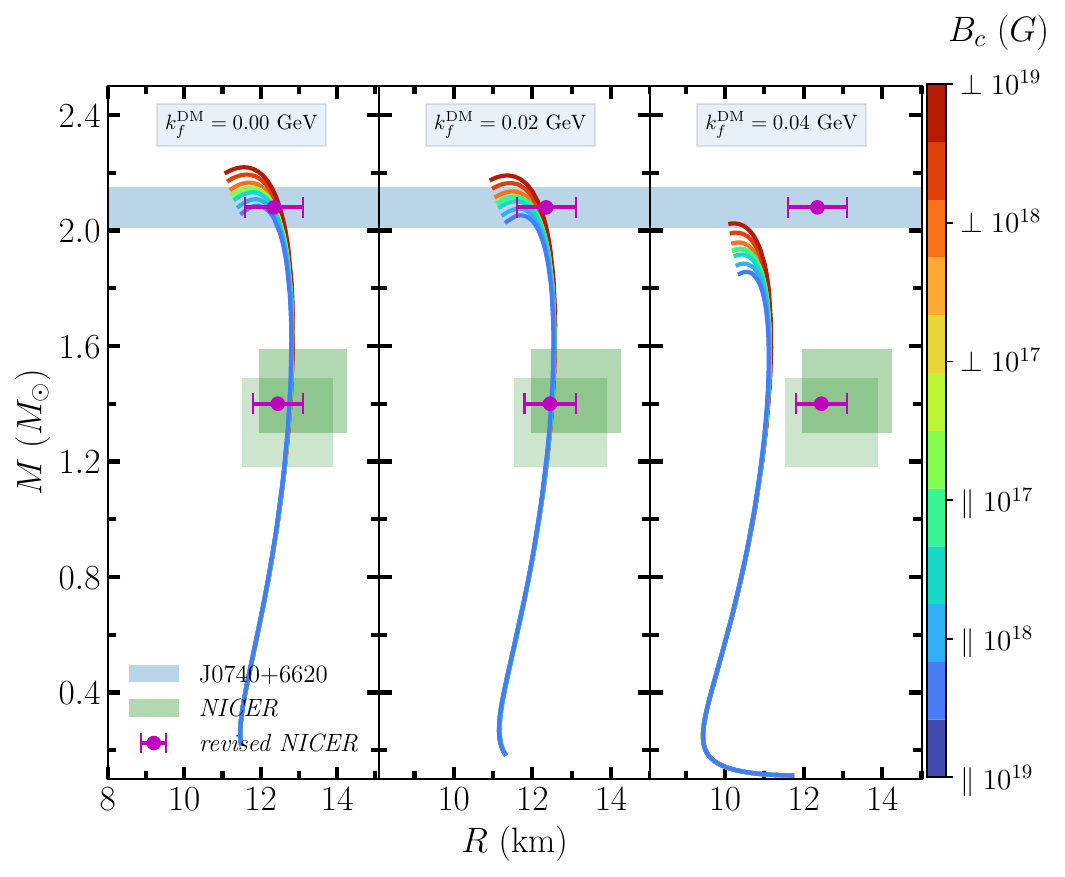}
    \caption{Mass-Radius relations for the DM admixed magnetized neutron star with varying the MF strength for BigApple (left) and IOPB-I (right). The overlaid bands are the observational data given by different observations (see text for details).}
    \label{fig:mr}
\end{figure*}
%%%%%%%%%%%%%%
%%%%%%%%%%%%%%%%%%%%%%%%%%%%%%%%%%%%%
\subsection{ Particle Fractions and EoS}
%%%%%%%%%%%%%%%%%%%%%%%%%%%%%%%%%%%%%
The particle fraction (PF) of the species, such as neutrons, protons, electrons, muons, and DM, can be calculated using the formula 
\begin{equation}
    X_i = \rho_i/\rho_b \, ,
\end{equation}
%%%%%%%%%%%%%%
where $\rho_i$ is the density of each species, and $\rho_b$ is the baryon density. 

In Fig. \ref{fig:eos_par_frac}, we calculate the value of $X_i$ for DM admixed magnetized neutron star for three different DM momenta $k_f^{\rm DM} = 0.00, 0.02$, and $0.04$ GeV with the variation of the core MF. The PF for neutron and DM are not taken as they are charge-less particles. Hence, the MF has no interactions with them. In this study, we have not included the anomalous magnetic moment (AMM) of the neutrons and protons for simplicity of the results. 

From the upper panel of Fig. \ref{fig:eos_par_frac}, we notice that protons and electrons appear almost at the same density. However, the muon appears $\approx 0.1$ fm$^{-3}$ for both BigApple and IOPB-I cases. It is observed that a lower magnitude of MF (for example, $10^{17}$ G) doesn't change the PF significantly. However, with an increase in the MF strength at the core, the population density changes and shows oscillating behavior due to the subsequent filling of Landau levels. 
This is because, with an increase in MF strength, the mass of the charged particle becomes heavier, and the oscillatory behaviour becomes prominent, especially in the core of the neutron star, as mentioned in Ref. \cite{Patra_2020}. 
In addition to MF, it is also noticed that the neutron star with a finite DM fraction does not affect the PF.

The EOSs for magnetized neutron stars with DM admixture are calculated and presented in the lower panel of Figure \ref{fig:eos_par_frac}, illustrating the dependence on the MF strength for various DM fractions. The EOSs exhibit increased stiffness (or softness) for the $\perp$ ($\parallel$) pressure.
While the   $P_\perp$ and   $P_\parallel$ differ in their magnitude (see Eqs. \eqref{eq:p_par} and \eqref{eq:trans_press} ); they only become significantly different for very large $B$ \cite{Strickland_2012}. 
Therefore, the anisotropy in the pressure becomes significant only at the higher strength of the central magnetic field. Furthermore, the difference between the $\perp$ and $\parallel$ pressure (stiffness ( softness) of the $\perp$ ($\parallel$) pressure) primarily arises due to the sign (+ (-)) of the contribution of the pure magnetic term in the $\perp$ ($\parallel$) pressure as in Eq. \eqref{eq:press} (also see Fig. 6 and the associated description of Ref. \cite{Mariani_2022}). Additionally, the presence of DM introduces further softening effect on the EOSs, with the degree of softness primarily determined by the DM content within the neutron star. Comparing the models employed in this study, BigApple demonstrates a stiffer EOS than IOPB-I. However, the influence of the MF on the softness or stiffness of the BigApple model is relatively less prominent as compared to the IOPB-I case. This model dependency of the MF effect on EoS stems from their effective masses which control the number of Landau levels (see Eq. \eqref{eq:eff_mass} and \eqref{eq:nmax}). 

In a previous study \cite{Parmar_2023}, we computed the equation of states (EOSs) for magnetized crusts using the CLDM model. In the present investigation, we employ the same model to determine the crust EOS. However, for the core EOS, we consider robust mean-field (RMF) models, namely BigApple and IOPB-I. Subsequently, we construct unified EOSs encompassing the BigApple and IOPB-I cases, as depicted in Figure \ref{fig:eos_par_frac}. It is worth noting that the percentage of DM remains nearly constant throughout the neutron star, predominantly concentrated within the core region  (single-fluid model)\cite{Panotopoulos_2017, Das_2019}.  The MF, on the other hand, exerts negligible influence on the crust EOS. Consequently, the lower-density EOSs for both the BigApple and IOPB-I models exhibit minimal variation as a function of  MF strength and DM fractions.

%%%%%%%%%%%%%%%%%%%%%%%%%%%%%%%%%%%%%%%%%%%%%%%%%%%%%%%
\subsection{Mass-Radius relations, tidal deformability}
%%%%%%%%%%%%%%%%%%%%%%%%%%%%%%%%%%%%%%%%%%%%%%%%%%%%%%%

The mass-radius ($M-R$) relations are obtained with Tolmann-Oppenheimer-Volkoff \cite{TOV1, TOV2} equations for the range of central densities, which is shown in Fig. \ref{fig:mr}. We calculate the $M-R$ profiles for both BigApple and IOPB-I EOSs with $\parallel$ and $\perp$ components of pressure (taking $P_\parallel$ and $P_\perp$ in the TOV equations) by varying DM momenta 0.00, 0.02, and 0.04 GeV. The magnitude of the maximum mass and its corresponding radius decreases for $P_\parallel$ and vice-versa for the $P_\perp$. In addition to MF, the DM also reduces the magnitude of $M$ and $R$ values, which depend on its percentage inside the star. In the case of the BigApple (IOPB-I) case, the maximum mass is $2.60 \ (2.15) M_\odot$, and its corresponding radius is $12.41 \, (11.91)$ km without the inclusion of the DM and MF.

With the inclusion of MF/DM, the magnitude of the maximum mass and its corresponding radius decreases $\sim 4-5 \%$. For $k_f^{\rm DM}=0.00, 0.02$ GeV with all MF components, the curves corresponding to the BigApple model reasonably satisfy the overlaid observational data. However, with $k_f^{\rm DM} = 0.04$ GeV, all curves satisfy only the maximum mass constraint given by PSR J0740+6620 \cite{Cromartie_2020}. Moreover, the canonical radii corresponding to 0.04 GeV for the BigApple case don't pass through NICER and revised NICER limits \cite{Miller_2019, Miller_2021}. In the case of the IOPB-I parameter set (right panel of Fig. \ref{fig:mr}) with $k_f^{\rm DM}=0.00, 0.02$ GeV, only parallel components of pressure  satisfy all the constraints imposed by Cromartie {\it et al.}, pulsar, and NICER. However, for $k_f^{\rm DM}=0.04$ GeV, none of the constraints is satisfied for IOPB-I set. Hence, from this study, we observe that one can put constraints on the amount of DM percentage and the strength of the MF  by employing diverse observational data.  It is relevant to mention here that  the information on the mass-radius profile of the neutron star essentially comes from gravitational waves (GW), pulsar hotspot, and x-ray measurements. Observational data from these techniques regarding the mass and radius of neutron stars typically do not directly account for the presence of  magnetic field. However, the majority of the information on the mass and radius of the star  comes from the pulsars, which are rotating neutron stars with surface magnetic field strengths of the order of $10^{10-15}$ G. Although various magnetic properties of these objects are measured, such as surface magnetic field, magnetic dipole moment, etc., their mass-radius profile is not mapped to the strength of the magnetic field. In other words, we do not yet have observational evidence of the influence of the magnetic field on the mass-radius profile of the star, primarily due to the unknown nature of the neutron star's inner core.

%%%%%%%%%%%%%%%%%%%%%%%%%%%%%%%
\begin{figure}
    \centering
    \includegraphics[width=0.55\textwidth]{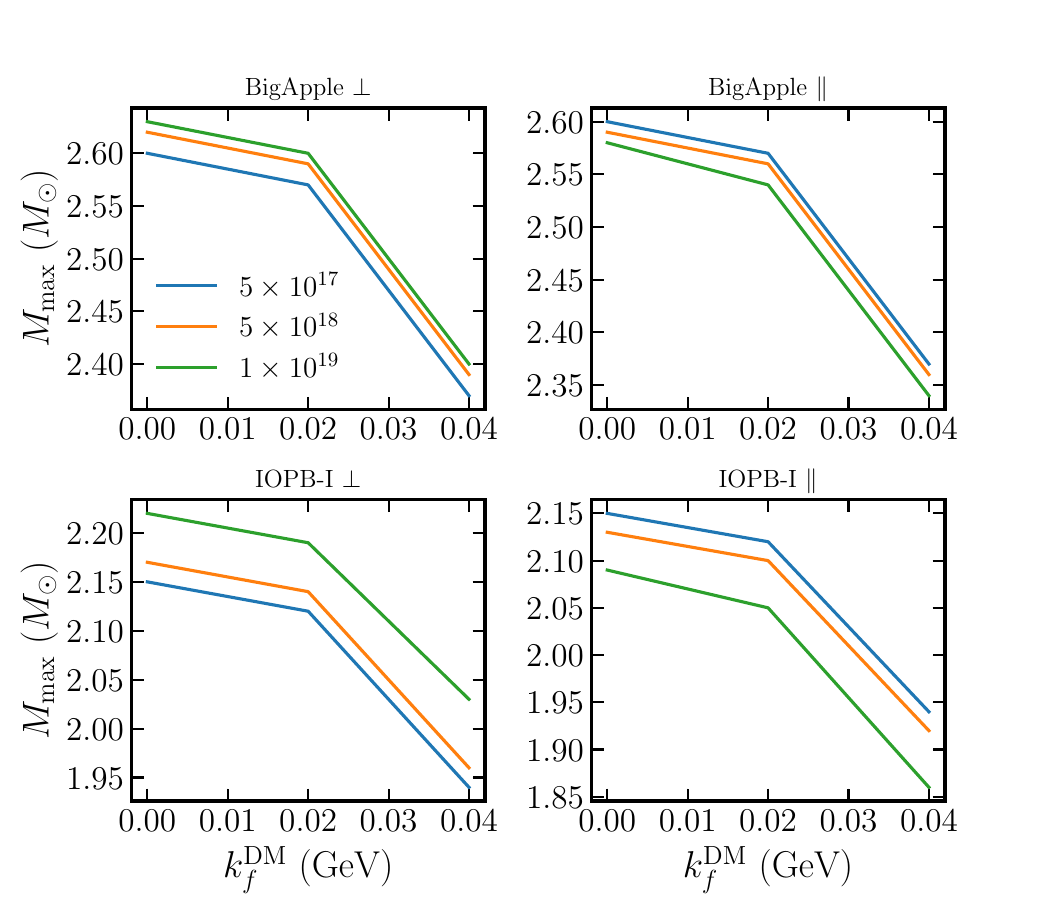}
    \caption{Maximum mass of magnetized neutron star admixed with the DM as a function of DM momentum for different MF values.}
    \label{fig:mr_rate}
\end{figure}
%%%%%%%%%%%%%%%%%%%%%%%%%%%%%%

Through the preceding analysis, it is evident that the EOS and mass-radius profile of DM admixed magnetized neutron stars are governed by two competing mechanisms. Firstly, the EOS experiences stiffening (softening) due to the $\perp$ ($\parallel$) pressure, while secondly, it undergoes softening as a consequence of increased DM content. To investigate the potential influence of MF on the rate at which the neutron star mass decreases due to the presence of DM, we construct a plot of the maximum mass as a function of DM Fermi momenta for varying MF strengths, as illustrated in Fig. \ref{fig:mr_rate}. Remarkably, for higher MF strengths, the rate of maximum mass reduction caused by DM content exhibits a slight decrease. In other words, the MF exerts an attenuating effect on the DM within the neutron star. Nonetheless, this effect is primarily notable under high MF strengths, indicating that the MF does not significantly impact the DM's influence on the properties of neutron stars. 
%%%%%%%%%%%%%%%
\begin{figure*}
    \centering
    \includegraphics[width=0.55\textwidth]{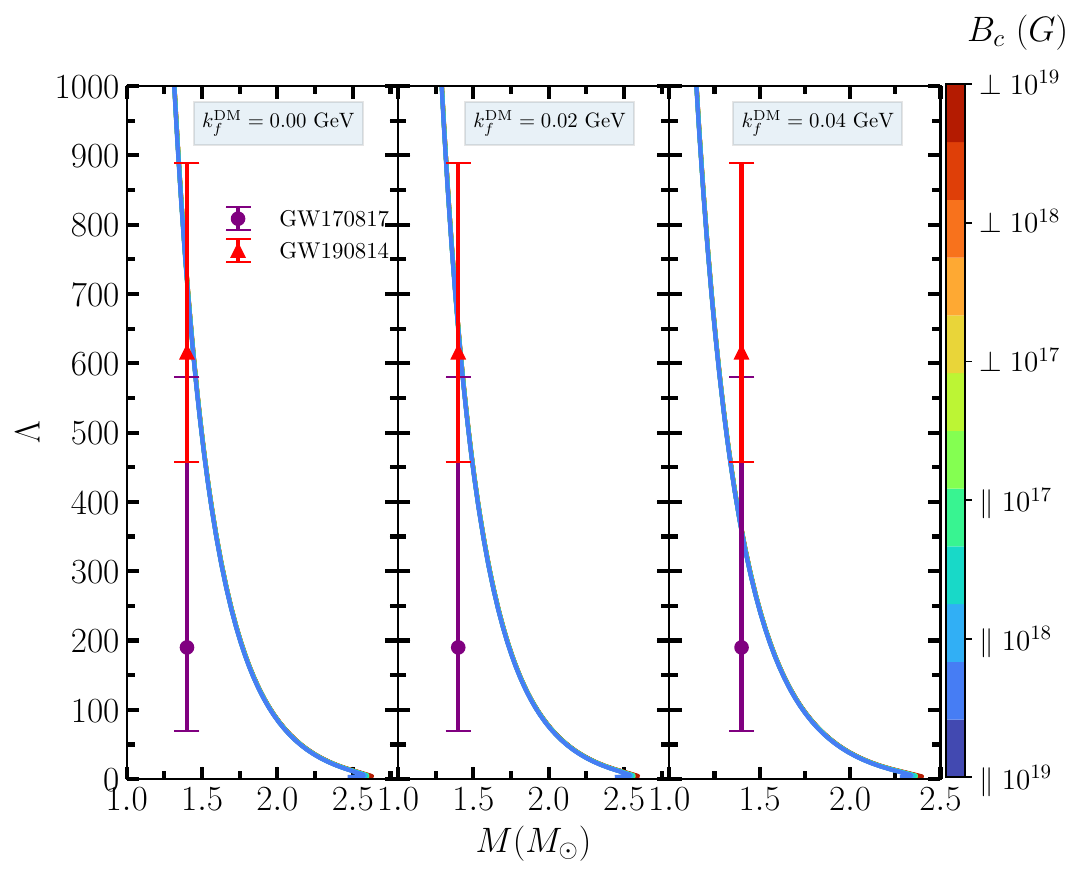}%%
    \includegraphics[width=0.55\textwidth]{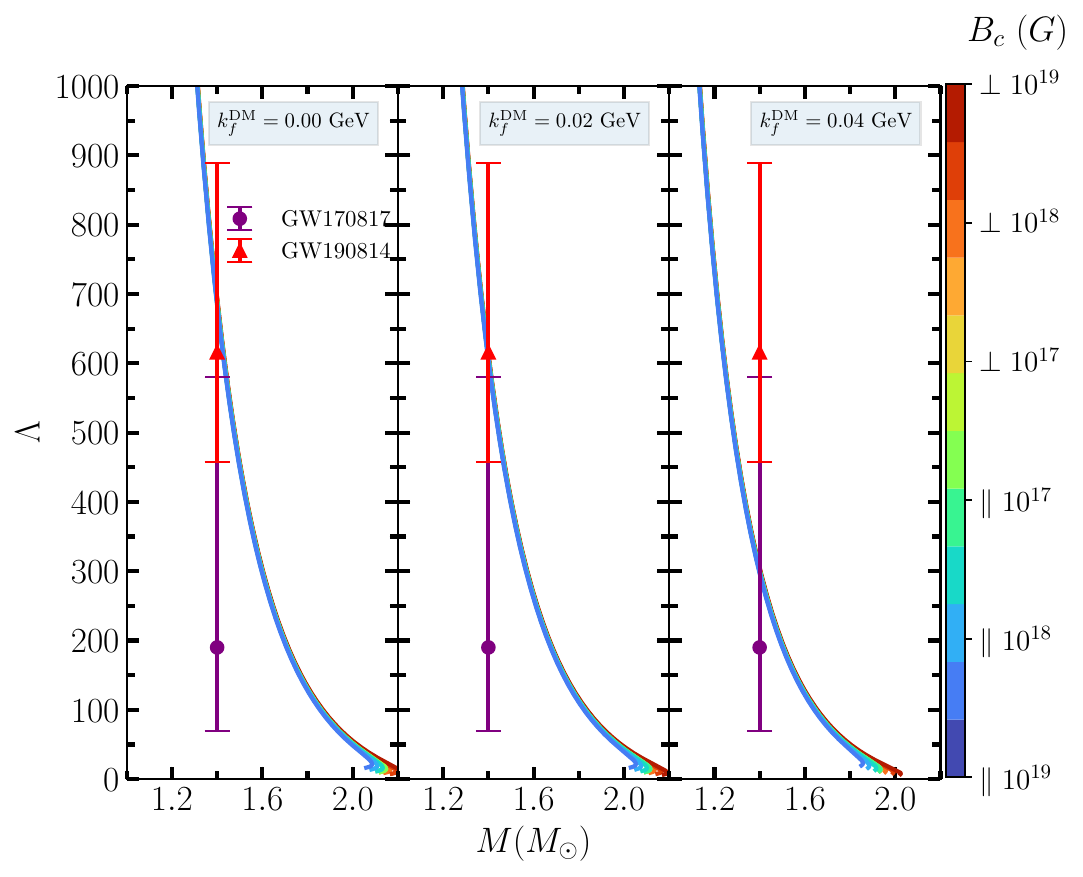}
    \caption{Dimensionless tidal deformability for the DM admixed magnetized neutron star with varying the MF strength for BigApple (left) and IOPB-I (right). The value of $\Lambda_{1.4} = 190_{-120}^{+390}$, and $\Lambda_{1.4} = 616_{-158}^{+273}$ taken from the GW170817, and GW190814  events respectively.}
    \label{fig:tidal}
\end{figure*}
%%%%%%%%%%%%%%

Next, we calculate the dimensionless tidal deformability of the DM admixed magnetized neutron star. The dimensionless tidal deformability ($\Lambda$) of the star is calculated using the relation \cite{Hinderer_2008}
%%%%%%%%%%%%%%%%%%
\begin{equation}
    \Lambda = \frac{2}{3}k_2C^5 \, ,
\end{equation}
%%%%%%%%%%%%%%%%%%
where $k_2$ is the Love number for the quadrupole case. The solution for $k_2$ can be found in Refs. \cite{Hinderer_2008, DasBig_2020}. $C$ is the compactness defined as $M/R$. We calculate the $\Lambda$ for different DM momenta with the variation of MF, which is shown in Fig. \ref{fig:tidal} for BigApple and IOPB-I E-RMF sets. With the addition of DM, the values of $\Lambda$ decrease. The curves corresponding to all the magnetized EOSs, including different DM momenta, are observed to be relatively diminished in the lower mass regimes.
However, substantial changes have been observed at the maximum mass limit. Different error bars are taken from the GW170817 and GW190814 events to constrain the value of $\Lambda$. Except for DM momenta 0.04 GeV, none of the curves passes through the GW170817 data. However, all the curves well reproduced the GW190814 limit except for $k_f^{DM}=0.04$ GeV for both BigApple and IOPB-I cases. In comparison to DM, the change in the value of $\Lambda$ for both $\parallel$ and $\perp$ pressure components are bleak for the canonical mass (1.4 $M_\odot$), i.e.  the theoretical estimates of
$\Lambda_{1.4}$ are not sensitive to the presence of a magnetic field for both EOSs. However,  the $\Lambda$ changes considerably for maximum neutron star mass depending on the strength of MF. Hence, we observed a significant effect due to the DM for both BigApple and IOPB-I cases.

%%%%%%%%%%%%%%%
\begin{figure*}
    \centering
    \includegraphics[width=0.55\textwidth]{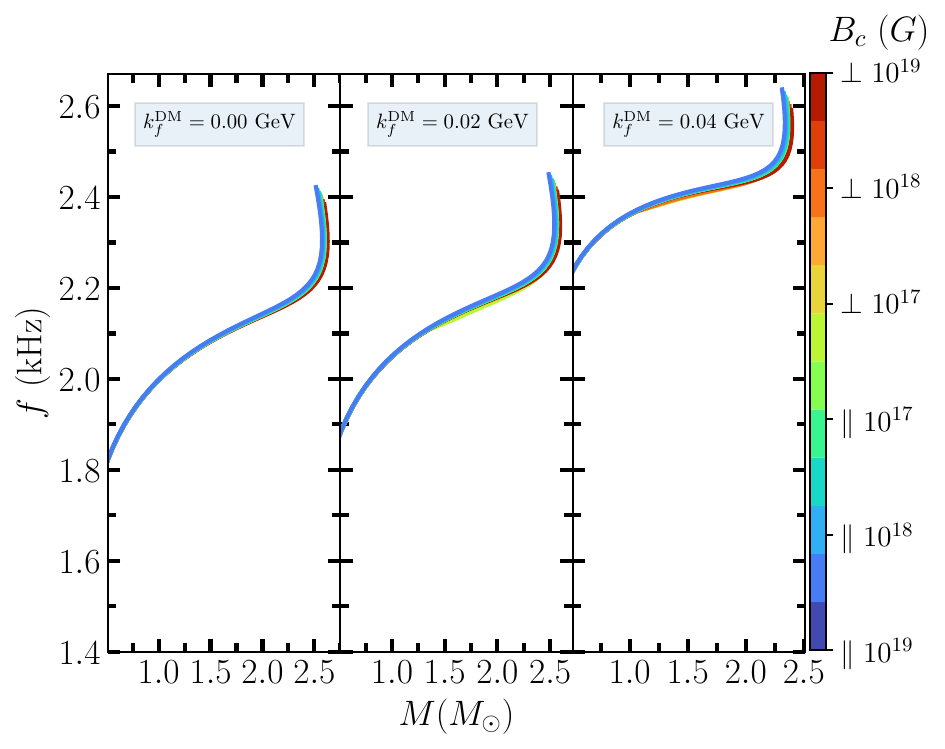}%%
    \includegraphics[width=0.55\textwidth]{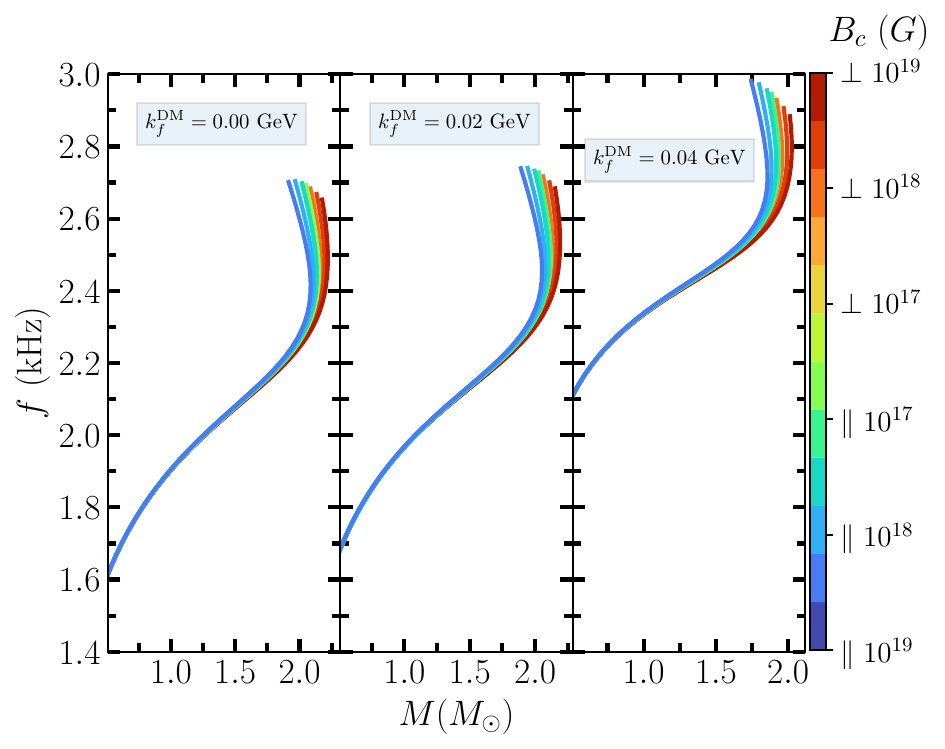}
    \caption{$f$-mode oscillation frequency of the magnetized neutron star with different fractions of DM for BigApple (left) and IOPB-I (right) cases.}
    \label{fig:fmode}
\end{figure*}
%%%%%%%%%%%%%%
%%%%%%%%%%%%%%%%%%%%%%%%%%%%%%%%%%%%%%%%%%%%%%%%%%%%%%%%%%%%%%%%%%%%%
\subsection{Calculation of $f$-mode oscillation of the magnetized neutron star}
%%%%%%%%%%%%%%%%%%%%%%%%%%%%%%%%%%%%%%%%%%%%%%%%%%%%%%%%%%%%%%%%%%%%%
In the present section, we use the formalism required to calculate the $f$-mode frequency as done in our previous study with the relativistic Cowling approximation \cite{Das_fmode_PRD_2021, flores2020gravitational}. The $f$-mode frequency for the quadrupole case is calculated with the variation of MF strength for different DM fractions, as shown in Fig. \ref{fig:fmode}. We find the marginal changes in the $f$-mode frequency in the case of the BigApple. With the increase in $k_f^{\rm DM}$, the EOS becomes softer, which gives the lower values of the maximum mass and its corresponding radius. Also, a relatively lower massive star oscillates with a higher frequency and vice-versa. Therefore, the magnitude of $f$-mode frequency for $k_f^{\rm DM} = 0.04$ GeV is higher than other DM momenta. We observe that the parallel and perpendicular MF components have marginal effects on the $f$-mode frequency of the star.  

In the case of IOPB-I (right panel in Fig. \ref{fig:fmode}), there are significant changes for both MF as well as DM. This is because the MF considerably affects the EOSs for the IOPB-I. The magnitude of $f$-mode frequency is higher for IOPB-I because it has a softer EOS compared to BigApple. Therefore, it oscillates with a higher magnitude and radiates more $f$-mode frequency.

%%%%%%%%%%%%%%%%%%%%%%%%%%%%%%%%%%%%%%%%%%%%%%%%%%%%%%%%%%%%%%%
\subsection{Relative change in the magnitude of neutron star properties}
%%%%%%%%%%%%%%%%%%%%%%%%%%%%%%%%%%%%%%%%%%%%%%%%%%%%%%%%%%%%%%%
%%%%%%%%%%%%%%%
\begin{figure*}
    \centering
    \includegraphics[width=0.8\textwidth]{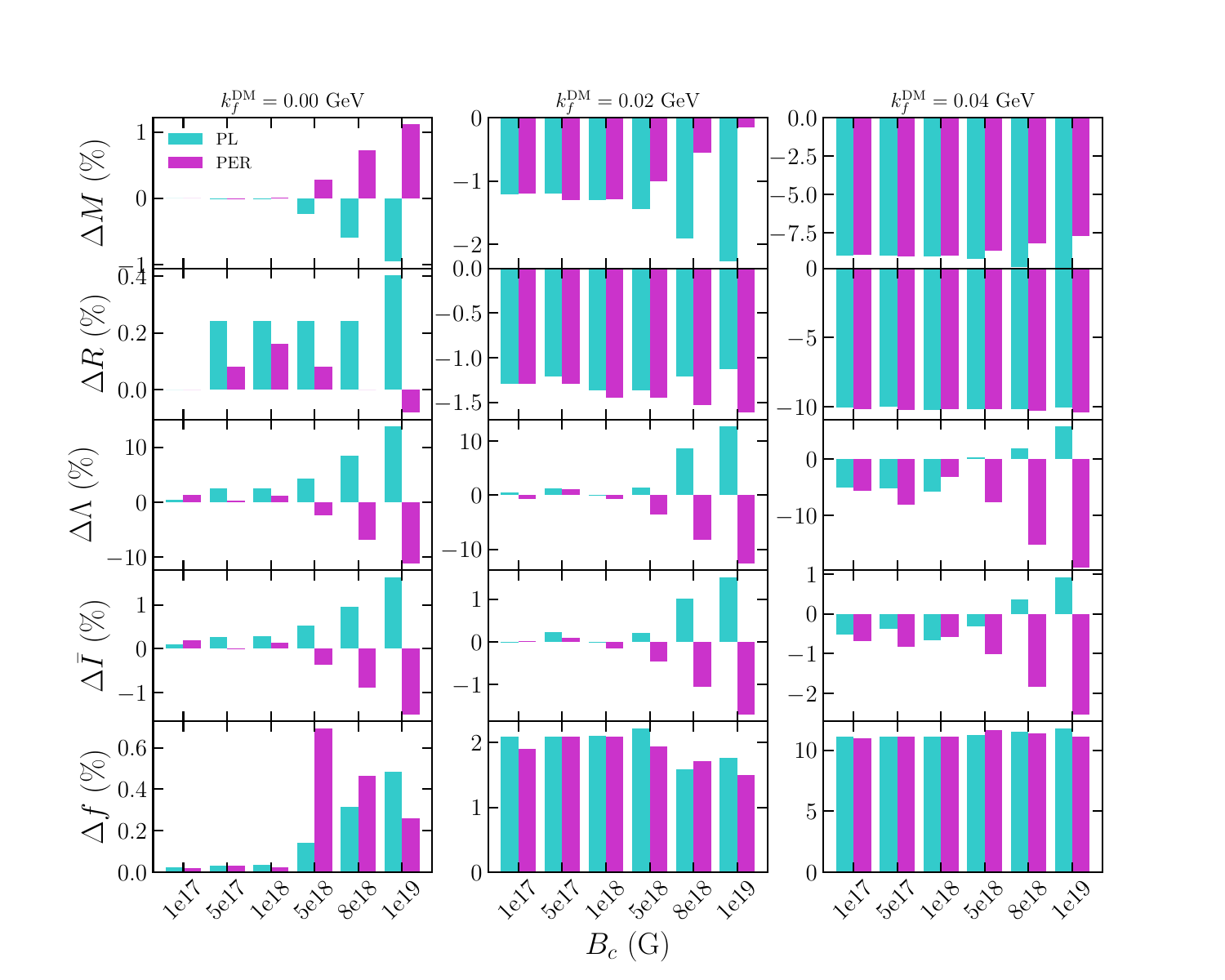}\\
    \includegraphics[width=0.8\textwidth]{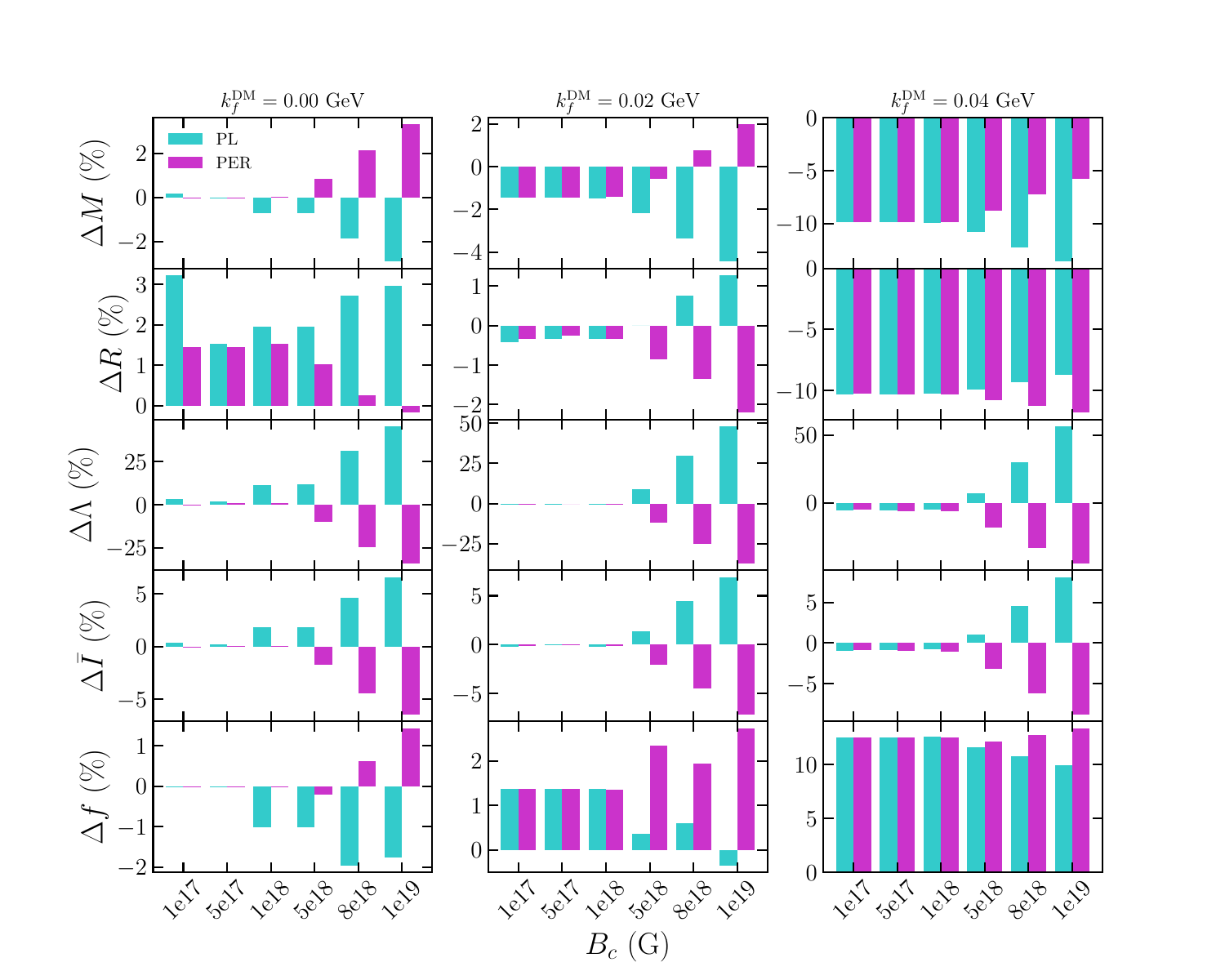}
    \caption{Relative change in the magnitude with respect to zero magnetic cases in the center ($B_c$) for $\Delta M$, $\Delta R$, $\Delta \Lambda$, $\Delta \Bar{I}$, and $\Delta f$-mode frequency compare with different DM fractions for BigApple (upper), and IOPB-I (lower) cases.}
    \label{fig:bar_change}
\end{figure*}
%%%%%%%%%%%%%%

Given the significant impact of both MF and DM on neutron star properties, it becomes imperative to thoroughly investigate and understand their relative influences on various aspects of neutron stars. To examine this, in  Fig. \ref{fig:bar_change}, we calculate the percentage change in the magnitude of the mass, radius, tidal deformability, dimensionless moment of inertia, and $f$-mode oscillation frequency in comparison to zero MF strength and DM content for three different DM percentage. The pink and cyan bars represent the perpendicular and parallel components of the pressure corresponding to that central MF ($B_c$).    

The relative changes in the neutron star properties are contingent upon three key components: (i) the strength of the MF, (ii) the percentage of DM, and (iii) the specific EOS employed. Notably, for MF strengths below $1 \times 10^{18}$ G, both the parallel and perpendicular pressure components exhibit negligible variations, as the emergence of anisotropy becomes evident only at higher MF strengths. In the presence of higher MF strengths, the parallel pressure component facilitates a decrease in the maximum mass due to the influence of DM, while the perpendicular pressure component mitigates the rate at which this decrement occurs. In contrast to the maximum mass, the radius of the star exhibits the opposite behavior. Specifically, for higher MF strengths, the perpendicular component of the pressure contributes to a decrease in the radius, while the parallel pressure component attenuates this effect. The maximum mass and corresponding radius of the star can exhibit changes as high as 10\%, highlighting significant variations in these properties due to the presence of MF and DM.

The dimensionless tidal deformability ($\Lambda$) and the normalized moment of inertia ($\Bar{I}$) are notably more influenced by the presence of MF compared to DM. Additionally, the changes in these properties demonstrate sensitivity to the chosen EOS. The combined effects of DM and MF offer a range of possibilities that prove valuable in reproducing specific observational data associated with $\Lambda$ and $\Bar{I}$.

Furthermore, the $f$-mode oscillation frequency is found to be influenced by the presence of MF and DM. The introduction of DM significantly increases the $f$-mode frequency, reaching up to 2\% for $k_f^{\rm DM}=0.02$ and 10\% for $k_f^{\rm DM}=0.04$. The presence of MF further impacts this frequency increase. The influence of MF on top of DM depends on the strength of the MF and the chosen EOS used in the analysis. 

\section{Summary and Conclusion}
\label{conclusion} 
%%%%%%%%%%%%%%%%%%%%
This study explores the different properties of the DM admixed magnetized neutron star. The magnetized EOSs are calculated with the relativistic mean-field model with density-dependent MF. The well-known RMF models, namely BigApple, and IOPB-I, are used to obtain the EOSs for the magnetized neutron star. In the case of the DM, we choose the simple DM model, where the DM particle interacts with nucleons by exchanging Higgs. The MF strength in the core is varied by fixing the surface MF. Moreover, the DM fraction is almost constant throughout the neutron star. To see its effect on the magnetized neutron star, we vary its fraction from $0.00-0.04$ GeV. We have calculated various properties such as mass, radius, tidal deformability, $f$-mode frequency with different interaction strengths for MF, and percentage of DM.

The EOSs of the DM admixed magnetized neutron star are found to be softer for the parallel component of the pressure and stiffer  for the perpendicular one. The softness also depends on the DM contained inside the neutron star. The particle fraction for the magnetized neutron star exhibits an oscillating nature (predominantly in the core region of the star), due to an increase in the proton mass and filling of Landau levels for the high MF. The macroscopic properties are significantly affected by the DM as well as the MF. A higher DM percentage having high MF strength in the core has pronounced effects on the neutron star.

It has been observed that the maximum mass and its corresponding radius decreases $\sim 4-5 \%$ for DM admixed neutron star. The resulting maximum mass and radius of the magnetized neutron star admixed with DM then result from the competitive behavior of MF strength and DM percentage. Furthermore, the rate at which the maximum mass decreases with increasing DM percentage is attenuated by the presence of MF. The combined inclusion of both DM and MF proves instrumental in reproducing certain observational data that could not be predicted in the absence of these interactions.

The tidal deformability ($\Lambda$) and $f$-mode oscillation frequency are obtained for both BigApple and IOPB-I cases by varying the MF strength and DM percentage. The $\parallel$ components have a higher magnitude of $\Lambda$ in comparison to $\perp$ one. We found similar results for the $f$-mode frequency of the star.  It has been observed that a significant change arises in ($\Lambda$) and $f$-mode oscillation frequency due to both MF and DM. 

In future studies, it is possible to explore additional macroscopic properties of both static and rotating DM admixed magnetized neutron stars. In this  work, a spherically symmetric neutron star model is considered as a simplifying assumption. However, to calculate the properties of neutron stars more efficiently, advanced techniques such as Lorene and numerical relativity can be employed for both the static and rotating cases. Such techniques may provide a comprehensive insight into the properties of DM admixed magnetized neutron stars.
%%%%%%%%%%%%%%%%%%%%%%%%%%%%%
\section{Acknowledgement}
We would like to thank the anonymous reviewers for the careful reading of our manuscript and their
insightful suggestions. We sincerely appreciate all valuable comments and suggestions, which helped us to improve the quality of the manuscript. 

%%%%%%%%%%%%%%%%%%%%%%%%%%%%%
\bibliography{dmmf}
\bibliographystyle{apsrev4-2}
%%%%%%%%%%%%%%%%%%%%%%%%%%%%%
\end{document}